\documentclass[11pt]{article}
\usepackage[a4paper, top=2.0cm, bottom=2.0cm, left=2.0cm, right=2.0cm]{geometry}
\usepackage{multicol}
\usepackage{setspace}
\usepackage[scaled]{helvet}
 
\usepackage[T1]{fontenc}
\usepackage{graphicx}
\usepackage{amssymb}
\usepackage{epstopdf}
\usepackage{wrapfig}
\usepackage{color}
\usepackage{chemformula}
\usepackage{caption}
\usepackage{upgreek}
\usepackage[dvipsnames]{xcolor}
\usepackage[font={color=black},figurename=Fig.,labelfont={it}]{caption}
\usepackage{siunitx}
\usepackage{authblk}
\usepackage{hyperref}
\usepackage{xcolor}
\usepackage{enumitem} 
\usepackage{paralist}
\setitemize{noitemsep,topsep=0pt,parsep=0pt,partopsep=0pt,leftmargin=1.5\parindent}
\hypersetup{
colorlinks=true,
linkcolor=blue,
urlcolor=blue,
citecolor=blue,
}
\usepackage{natbib}
\usepackage{aasmacros}
\usepackage{doi}
\usepackage[labelformat=empty]{caption}
\usepackage{soul}

\newcommand{\squishlist}{
 \begin{list}{$\bullet$}
  { \setlength{\itemsep}{1pt}
     \setlength{\parsep}{0pt}
     \setlength{\topsep}{1pt}
     \setlength{\partopsep}{0pt}
     \setlength{\leftmargin}{1.5em}
     \setlength{\labelwidth}{1.5em}
     \setlength{\labelsep}{0.5em} } }
\newcommand{\squishend}{
  \end{list}  }

\begin{document} 

\title{Transiting exoplanets as the immediate future for population-level atmospheric science \textit{(thematic area: Astro)}}

\author[1]{\textbf{Joanna K. Barstow}*}
\author[2]{\textbf{Hannah R. Wakeford}*}
\author[3]{Sarah L. Casewell}
\author[4]{Vatsal Panwar}
\affil[1]{School of Physical Sciences, The Open University, Walton Hall, Milton Keynes, MK7 6AA}
\affil[2]{School of Physics, HH Wills Laboratory, Tyndall Avenue, Bristol, BS8 1TL}
\affil[3]{Department of Physics and Astronomy, University of Leicester, University Road, Leicester, LE1 7RH}
\affil[4]{School of Physics \& Astronomy, University of Birmingham, Edgbaston, Birmingham, B15 2TT}

\affil[*]{Contact: Jo.Barstow@open.ac.uk, hannah.wakeford@bristol.ac.uk}

\date{}

\maketitle

\section{Scientific Motivation \& Objectives}

The transit method, during which a planet's presence is inferred by measuring the reduction in flux as it passes in front of its parent star, has become the most successful exoplanet detection technique over the last several decades. A major advantage of this technique is the opportunity it affords to not only detect a planet, but also characterize its atmosphere. During transit, the small fraction of starlight that passes through a planet's atmosphere emerges with the fingerprints of atmospheric gases, aerosols and structure. On the other side of the orbit, during eclipse, the relatively small contribution of thermal emission and reflected light from the planet itself may be observed as the planet is occulted by the star. For planets in near-edge-on, short-period orbits around their parent stars, observing the system throughout an entire orbit allows the varying flux from the planet to be extracted as the illuminated `dayside' rotates in and out of view. These close-in planets, often on periods less than 10 days, are tidally locked to their stars such that their rotation matches their orbital period, meaning they have a permanent dayside facing their stellar host and permanent nightside facing out into space. Spectroscopic observations across these observing methods provide an opportunity to characterize not only the overall composition of the atmosphere, but also glean insights into atmospheric structure and dynamics. 
\begin{wrapfigure}[16]{r}{0.47\textwidth}
  \begin{center}
    \includegraphics[width=0.45\textwidth]{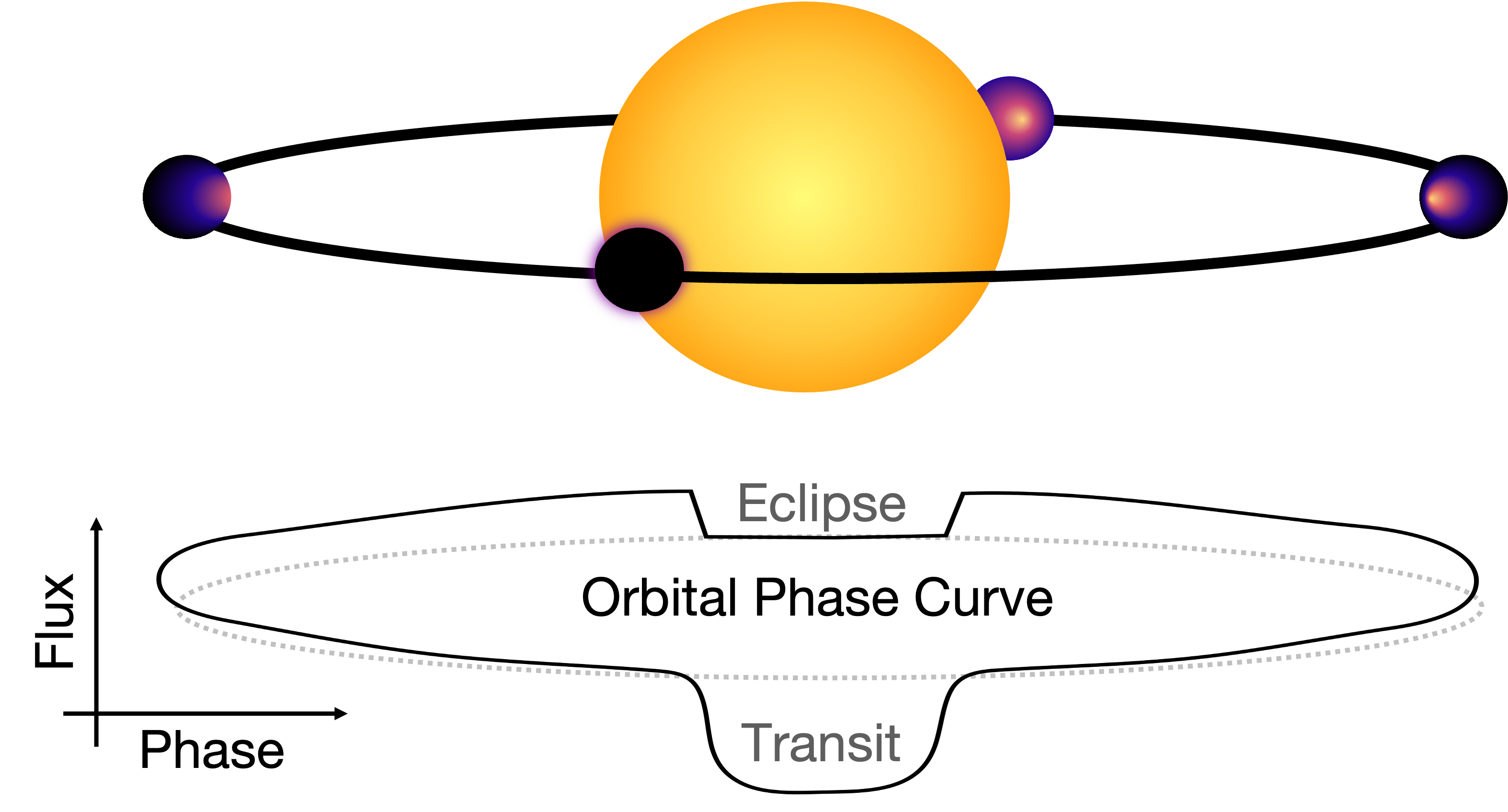}
  \end{center}
  \caption{\textit{Figure 1: Illustration of the transit method and the measured deviation in flux caused by the transit, eclipse, and phase curve of a tidally locked planet.}}
      \label{fig:transiting_planets}
\end{wrapfigure}

Whilst the transiting sample is biased towards larger, hotter planets in close orbits around their parent stars, objects with sub-Neptune masses and equilibrium temperatures of just a few hundred Kelvin are becoming accessible to transit spectroscopy, currently with \textit{JWST} and soon with \textit{Ariel}. Technology that will allow the study of mature planets at wide separations (thus non-tidally locked), including solar system analog planets, is currently in development (e.g., ELTs, the Habitable Worlds Observatory and LIFE missions) but will not be operational until the 2030s from the ground and 2040s from space. With over 6,000 transiting planets now discovered across a range of sizes, masses and orbiting stars of different spectral types, transit, eclipse and phase curve observations are currently our only window into a consistent sample of planetary atmospheres that is large enough to attempt a population-level study \citep[e.g.,][]{sing2016,Mansfield2021NatAs,Tinetti2022,valentine2025}. 

Population-level statistics are a critical next step for our understanding of atmospheric science. The vast exoplanet population provides a laboratory for atmospheric physics, including chemistry, dynamics, cloud processes and evolution. JWST has already produced a number of groundbreaking results, including constraints on horizontal wind speeds and quenching \citep[e.g.,][]{bell2024}; internal heating \citep[e.g.,][]{sing2024,welbanks2024}; cloud composition \citep[e.g.,][]{grant2023}; atmospheric escape \citep[e.g., ][]{krishnamurthy2025_wasp107b_helium}; and photochemistry \citep[e.g.,][]{alderson2023,powell2024,Gressier2025AJ}. Extending these results to a larger number of targets (hundreds) will allow us to explore the effects of equilibrium temperature, gravity, mass and parent star type on atmospheric properties, as well as map observable trends to formation scenarios \citep[e.g.,][]{kirk2024}. Ultimately, atmospheric observations of the transiting exoplanet population allow us to better constrain and understand atmospheric processes and their response to different forcings, as well as constrain models of planet formation. \textbf{These findings are critical for addressing STFC's Science Vision Challenge B: \textit{how do stars and planetary systems develop and how do they support the existence of life?}.} Similar questions were also highlighted in the United States Astro2020 Decadal: \textit{What fundamental planetary parameters and processes determine the complexity of planetary atmospheres?} (E-Q2c), \textit{How does a planet's interaction with its host star and planetary system influence its atmospheric properties over all time scales?} (E-Q2d), and \textit{How do giant planets fit within a continuum of our understanding of all sub-stellar objects?} (E-Q2e). Here we outline how the UK must fit within this strategic context, provide proposed approaches for the future and the required development to implement those approaches, as well as outlining the unique capabilities and leadership of scientists across the UK.

 \section{Strategic Context}
We are now entering the era of population-level characterization of transiting exoplanets and of exoplanet demographics. A number of missions and observing campaigns over the last two decades have been focused on the detection of transiting planets, with various target sampling strategies. For example, the Kepler space observatory \citep{Borucki2016kepler} surveyed a single area of sky in depth over the 4 years of its nominal mission whilst the TESS mission \citep{Ricker2014TESS} has covered the whole sky detecting planets around the brightest host stars. Over the next few years, this sample will be extended by the PLATO mission \citep[][launching 2026]{barker2025plato}, designed to identify rocky planets at $\sim$ 1 au distances orbiting sun-like stars, and the Nancy Grace Roman Space Telescope (\textit{Roman}) which will discover as many as several 10,000s of planets via the \textit{Roman} galactic bulge survey \citep{Johnson2020_roman}. These detections will sit within the wider context of new discoveries of non-transiting planets detected via microlensing (\textit{Roman} is due to launch May 2027) and astrometry (\textit{Gaia} data release 4, due late 2026) for an unprecedented understanding of planet occurrence rates and system architectures. 

Many space-based targets and those from the numerous ground-based surveys (e.g., Super-WASP, HATNet, NGTS, KELT) are suitable for follow-up and atmospheric characterization. \textit{JWST} has hugely expanded our ability to characterize exoplanets in transit, with access to the infrared and increased sensitivity over \textit{Hubble} providing access to crucial molecular absorbers at longer wavelengths, and allowing us to push down to increasingly smaller classes of planet. The Ariel mission, due to launch in 2029 \citep{Tinetti2022}, aims to survey the atmospheres of 1,000 transiting exoplanets via transits, eclipses and phase curves, with spectroscopic coverage in the near- and mid-infrared. With the Hubble Space Telescope continuing to provide data for the UV and optical for bright targets.

Following on from \textit{Ariel} and \textit{JWST}, transiting exoplanet science will continue with the Habitable Worlds Observatory (\textit{HWO}) in the 2040s, this time with UV-optical-near infrared coverage. \textit{HWO} is currently in development and simulation work to identify appropriate instrumentation will be critical over the next few years. Whilst \textit{JWST} is hoped to be operational up to 20 years from launch, \textit{Ariel}'s nominal mission is 4 years, with the option to extend by another 2 years. \textit{Hubble}, currently our only route of space-based access to the UV and blue optical, has now been operating for 35 years, and it is 16 years since its last servicing mission in 2009. It is currently operating under restricted conditions with a single gyroscope to prolong its lifetime. \textbf{This means that there is likely to be a gap in transiting exoplanet characterization science from space in the late 2030s/early 2040s}, and considering \textit{Hubble}'s age we are likely to lose UV-optical coverage earlier than this (expected de-orbit in the mid-2030s). This will almost certainly mean a delay in space-based follow up for promising transiting targets identified by \textit{PLATO} and \textit{Roman} in the late 2020s/early 2030s, unless other platforms can fill the gap. 

\section{Proposed Approach}
 
An ideal platform for the study of transiting exoplanet atmospheres has broad, simultaneous wavelength coverage; moderate spectral resolving power (minimum, $R\sim$3000); a low noise floor ($\sim$10 ppm); and high stability over long periods of time. These are all critical because any measurement, whether transit, eclipse or a phase curve, is a difference measurement of the changing overall flux from the system, and the spectroscopic variation in this is on the order of a few 10s -- 100 ppm. To disentangle the absorption signatures from different gases, we need to be able to resolve the shapes of the bands. The stellar flux can vary over the star's rotation period as spots come in and out of view, and this has a spectroscopic component too, so measurements taken days, weeks or months apart may not have the same baseline, and stitching together non-contemporaneous chunks of spectrum can be challenging. Simultaneous coverage across a broad wavelength range minimises this difficulty. 

Broad wavelength coverage from the near-ultraviolet through to the mid-infared is important for capturing all of the key elements of an exoplanet spectrum. The near-ultraviolet contains information about atmospheric aerosols, forests of atomic metal lines, absorption by sulphur species and signatures of stellar surface features that need to be corrected for. When looking in eclipse, the UV provides information on the directly reflected light from the planets surface or atmosphere that is otherwise masked by the thermal emission of the planet which can emerge as short as 0.5\,$\upmu$m. Figure\,2 shows how the reflected light from an exoplanet dominates at the short wavelengths with thermal taking over at longer optical-IR wavelengths. 
\begin{figure}
    \centering
    \includegraphics[width=0.9\linewidth]{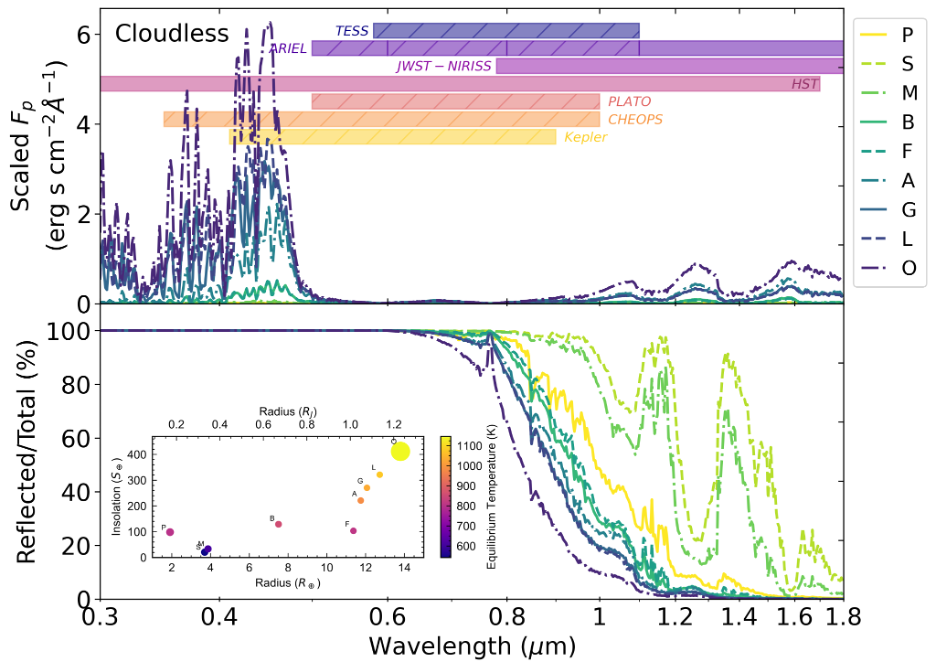}
    \caption{\textit{Figure 2: Top -- the total planetary flux as measured in flux units for cloud-free planetary atmospheres. 
    Bottom -- the percent contribution of reflected light to the total flux as a function of wavelength ($R$\,=\,300). Colored bands in the top panel show the photometric (hashed) and spectroscopic (solid) coverage from current and future facilities over this wavelength range. Figure updated from \citet{Mayorga2019AJ} and included with permission of L. C. Mayorga.}}
    \label{fig:mayorga2019}
\end{figure}
The optical is also important for aerosol scattering slopes, starspot signatures, atomic lines for key species such as sodium, and metal oxide/hydride features, as well as tracking atmospheric escape via the Hydrogen-alpha lines. The near-infrared contains a sequence of water vapour absorption features which are almost ubiquitous and help to set the baseline for transit observations, as well as features due to a vast range of other molecular species. Finally, the mid-infrared provides access to absorption features due to silicates that can reveal the composition of clouds. 

Observations of transiting exoplanets obtained in different geometries (in transit, in eclipse, or throughout the full phase curve) can reveal different information about the target. Transit spectroscopy is sensitive to relatively low atmospheric pressures of around 1 mbar (100 Pa) or less, since the long optical path length in transit geometry prevents light from penetrating through the deep atmosphere. It is also especially sensitive to scattering and extinction from aerosols and the overall atmospheric molecular weight, as well as abundances of specific gases.  By contrast, eclipse measurements probe deeper in the atmosphere in thermal emission, and are sensitive to the cloud tops for reflected light at short wavelengths. Thermal measurements are less sensitive to aerosols as cloud is more transparent in nadir geometry, but because eclipse measurements probe a broader range of pressures they are more sensitive to the vertical temperature structure. Although they probe different regions of the atmosphere, eclipse and transit measurements together provide a more holistic picture of the planet as a whole.

Phase curve measurements allow the mapping of thermal emission and reflected light around the planet as it completes an orbit of the star. This measurement technique is especially powerful because it can also be performed for non-transiting planets, provided they are close enough to edge-on for the change in flux as the illuminated dayside rotates in and out of view to be observable. Currently, phase curve observations are restricted to planets very close to their parent stars (orbital period $<2$ days), since they require observation throughout the orbital period and are therefore time-intensive. Longer-period phase curves would also require exceptionally high pointing stability from the telescope and instrument. Such measurements might become possible with \textit{HWO}. In the mean time, \textit{Ariel} could obtain phase curves of up to 40 planets over the mission lifetime, although with limited UV-optical coverage. 

\section{Proposed Technical Solution and Required Development}

For transiting exoplanet science to continue to flourish, continued investment in future space technology is required alongside funding to develop the facilities needed for interpreting space-based data. Such facilities include laboratory spaces for investigation into gas absorption data and cloud scattering properties, and to conduct fluid dynamics experiments. Currently, laboratory work in areas important for exoplanet atmospheric science may also fall within the remit of UKRI bodies apart from STFC (e.g., NERC or EPSRC) so it is crucial to have good communication between funding councils, as well as multidisciplinary calls for proposals to ensure that important science does not fall into gaps within existing funding structures. 

Also critical are high performance computing platforms such as DiRAC and IRIS, to allow for example simulations of planet formation, climate or interior structure; data reduction; calculation of energy levels for molecular absorbers; stellar surface models to allow for correction of starspot signatures, which includes more accurate MHD models for stellar photospheres; and the application of spectral retrieval tools to infer atmospheric solutions from observations. It also requires complementary observations from ground-based facilities, including high resolution spectra to measure planetary masses and potentially infer planet rotation rates and wind speeds, and long-term monitoring of the star to enable better correction for stellar activity. 

Alongside these supporting technologies, further investment in space instrumentation is a must. Longer-term, large missions such as NASA-led \textit{HWO} are likely to require hardware contributions from other nations, a need that the UK is very well placed to fill. Whilst NASA will take the lead on the workhorse coronagraph instrument, both a high resolution imaging instrument and an ultraviolet multi-object spectrograph are envisaged, either of which could be led by a UK team or have UK-manufactured components. 

Preparation for the immediate post-\textit{JWST}/\textit{Hubble}/\textit{Ariel} era is also necessary. Once operations cease for these telescopes, there is no currently planned space telescope that will be capable of broadband transiting exoplanet spectroscopy observations until the launch of \textit{HWO}. A particular need is UV coverage, which is likely to end soonest when \textit{Hubble} is no longer operational. It is critical to consider missions to fill this gap, especially since such observations will have significant value for planning observing campaigns with \textit{HWO}. 
Key to this goal are telescopes that can cover a broad wavelength range in the UV--optical (0.25--0.9\,$\upmu$m) with multi-colour photometry or  spectroscopic capabilities for transmission spectroscopy. This can even be achieved with a small aperture (20--60\,cm) telescopes with low-resolution spectroscopy providing an efficient way to make progress on providing population studies on well-characterised exoplanet systems, as well as providing applicable instrumentation to other fields of astronomy. 
Possibilities include WALTzER, a European Space Agency F-class candidate mission with a focus on UV spectroscopy of transiting exoplanets, Small Explorer Class missions through NASA such as exoplanet phase curve dedicated missions are in planning stages. 

What is clear is the critical need for information on these planets and their environments in the UV--IR over the coming decades, be it through physical laboratory based studies here on Earth, through to manifesting new telescopes to bridge the gap between current and planned observatories that will take us on the way to discovering and discerning planets like our own in the galaxy. 

\section{UK Leadership and Capability}

The UK has already displayed strong leadership in all of the key areas identified above. In hardware, with the UK at the forefront of UV detector development and imaging technology, and with two UK instrument feasibility studies currently funded and ongoing, UK provision of an additional instrument for \textit{HWO} either as lead or partner seems likely, budget permitting. This would help to cement the UK's role in the mission as well as providing economic benefit to UK-based companies. 

In terms of complementary science, the ExoMol group that calculates and hosts the ExoMol absorption line database is based in the UK. The UK exoplanet community also boasts expertise in ground- and space-based observation \citep[e.g.,][]{alderson2023,grant2023,vansluijs2023,Valentine2024AJ,sutlieff2025,gandhi2025}; aerosol modelling \citep[e.g.,][]{Gao2021JGRE..12606655G,Chubb2024MNRAS,lodge2024}; stellar surface modelling \citep[e.g.,][]{norris2023}; spectral retrieval and atmospheric inference \citep[e.g.,][]{macdonald2023,taylor2023,Fairman2024AJ,banerjee2024}; exoplanet climate simulations \citep[e.g.,][]{mak2024,barrier2025}; and exoplanet evolution \citep[e.g.,][]{booth2023MNRAS.518.1761B,owen2024MNRAS.528.1615O} and formation models \citep[e.g.,][]{booth2019MNRAS.487.3998B,penzlin2024}. The \textit{Ariel} mission is led from the UK and UK scientists win a high proportion of competitively awarded telescope time across space- and ground-based platforms; for example, in JWST Cycle 4 UK PIs led 13 accepted programs, the largest number for any individual country excluding the USA. 

Building on the experience of providing instrumentation for \textit{JWST} and leadership of \textit{Ariel}, a UK team would be well positioned to lead future F- and M- class missions for ESA targeting exoplanet science, which could continue the legacy of transiting exoplanet expertise after current and near-future missions are complete. 

\section{Suggested Mission Class}
An ESA mini-F, F- or M- class mission may be appropriate, depending on the exact science case definition. F-class missions are suitable for transiting exoplanet science, but for a more limited selection of potential targets compared with \textit{Ariel}, which is an M-class mission. 

As previously mentioned, WALTzER, is a proposed ESA F-class mission focusing on UV spectroscopy. Other key science goals can only be addressed in the IR such as measuring the thermal information from tidally locked exoplanets where full orbit phase curves are needed to explore trends in dynamics and formation. An example study on the use of orbital phase curves has been proposed for \textit{Ariel} but would nominally take 1--2 years to complete \citep{valentine2025}. An F- or M-class mission dedicated to phase curves could feasibly study $\sim$50 exoplanets in one year (given orbital periods of $<$5-days). Alternatively, a fleet (3 or more) of coordinated smaller space-based telescopes could be used to provide multi-band photometry and simultaneous coverage of longer period planets not feasible with \textit{Ariel}.

\section{Partnership Opportunities}
Many partnerships already exist through ESA (\textit{Ariel}) and NASA/Space Telescope Science Institute (\textit{JWST}, \textit{Hubble}). The UK's involvement in ESO is also critical for the continued success of UK transiting exoplanet science. Future collaboration with NASA on \textit{HWO} is likely, with the possibility of either leading, or joining with another nation/ESA in a bilateral agreement, to provide an instrument. 

\section*{Co-signatories}
\textbf{Cardiff University:}	
Lorenzo V. 	Mugnai,
Subhajit	Sarkar;
\textbf{Cornell University, University of Bristol	:}	
Lili	Alderson;
\textbf{Keele University:}	
John	Southworth;
\textbf{MSSL/UCL:}	
Vincent Van Eylen;
\textbf{Queen Mary University of London:}	
Matthew	Battley,
Edward	Gillen;
\textbf{RAL Space:}	
Chris	Pearson;
\textbf{The Open University:}	
David	Arnot,
Carole	Haswell,
Eleni	Tsiakaliari;
\textbf{University College London:}	
Jonathan	Tennyson,
Sergei N. 	Yurchenko;
\textbf{University of Bern:}	
Marrick	Braam,
Joost P. 	Wardenier;
\textbf{University of Birmingham:}	
Annelies	Mortier,
Adam	Stevenson,
Amaury	Triaud;
\textbf{University of Bristol:}	
Katy L. 	Chubb,
Charlotte	Fairman,
Daniel	Valentine;
\textbf{University of Cambridge:}	
Lalitha	Sairam;
\textbf{University of Chicago:}	
Dominic	Samra;
\textbf{University of Edinburgh:}	
Beth	Biller,
Trent	Dupuy,
Aiza	Kenzhebekova,
David A. 	Lewis,
Larissa	Palethorpe,
Paul	Palmer,
Mia Belle	Parkinson,
Ken	Rice,
Sarah	Rugheimer,
Ben	Sutlieff;
\textbf{University of Glasgow:}	
Matthew I. 	Swayne;
\textbf{University of Leeds:}	
Richard	Booth;
\textbf{University of Leicester:}	
Matthew	Burleigh;
\textbf{University of Oxford:}	
Suzanne	Aigrain,
Jayne	Birkby,
Claire	Guimond,
Tad	Komacek,
Mei Ting	Mak;
\textbf{University of St Andrews:}	
Ryan	MacDonald,
\textbf{University of Warwick:}	
David J. A. 	Brown,
Alastair	Claringbold	

\end{document}